\documentclass[journal=jpclcd,manuscript=article, layout=twocolumn]{achemso}

\usepackage[version=3]{mhchem} 
\usepackage{xcolor}
\usepackage{soul}

\author{Magnus A. H. Christiansen}
\email{mah107@hi.is}
\author{Alejandro Peña-Torres}
\author{Elvar Ö. Jónsson}
\author{Hannes Jónsson}
\email{hj@hi.is}
\affiliation[University of Iceland]
{Science Institute and Faculty of Physical Sciences, U. of Iceland, 107 Reykjavík, Iceland}

\title[An \textsf{achemso} demo]
  {Single-Atom Substituents in Copper Surfaces May Adsorb Multiple CO Molecules}

\begin{document}


\begin{tocentry}

\includegraphics[width = \linewidth]{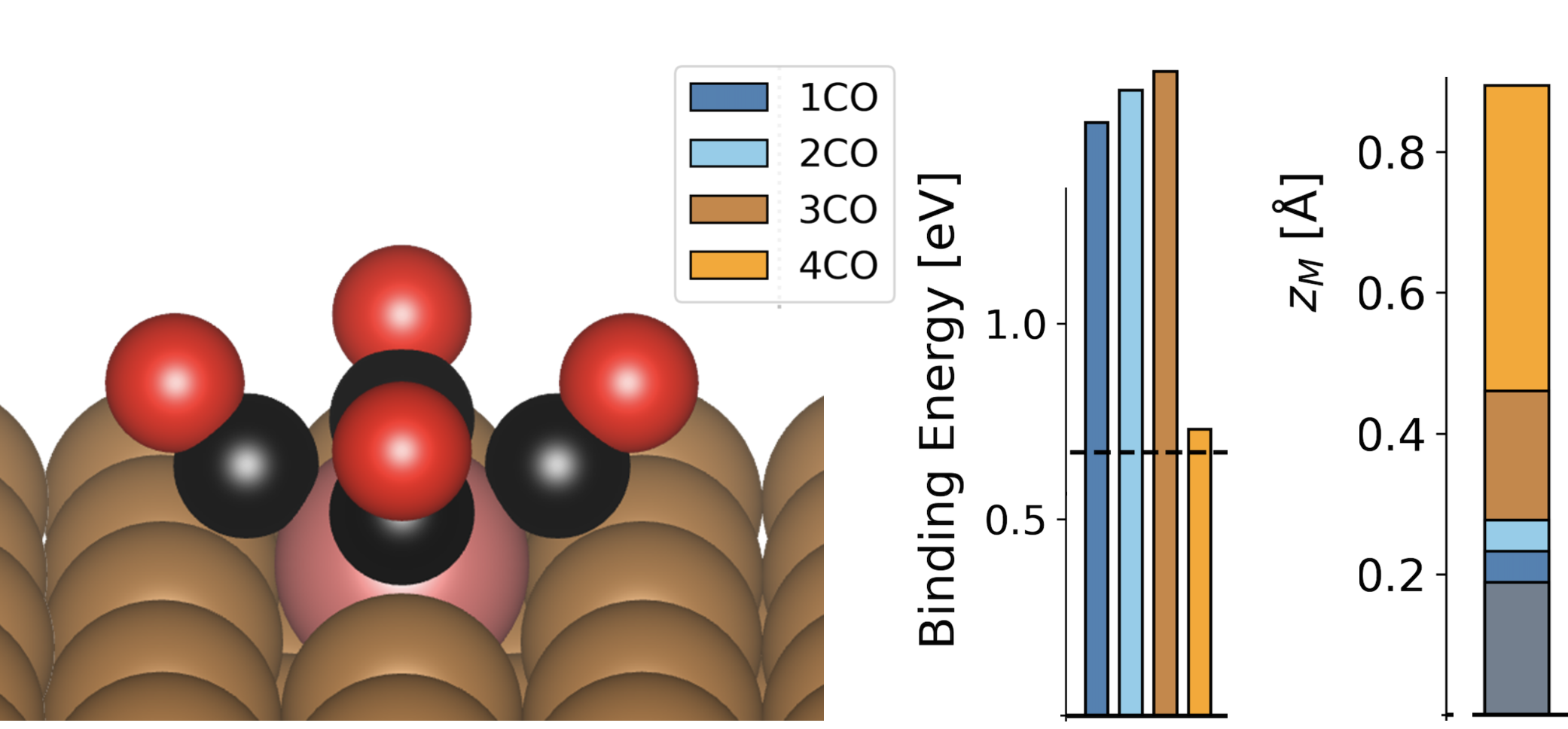}

\end{tocentry}

\begin{abstract}
Copper is a good CO$_2$ electroreduction catalyst as products beyond CO form, but efficiency and selectivity is low. Experiments have shown that an admixture of other elements can help, and computational screening studies have pointed out various promising candidates based on the adsorption of a single CO molecule as a descriptor. Our calculations of CO adsorption on surfaces where a first row transition metal atom replaces a Cu atom show that multiple CO molecules, not just one, bind to the substitutional atom. For Fe, Co, and Ni atoms, a decrease in binding energy is found, but the reverse trend, namely increasing bond strength, is found for V, Cr, and Mn and the first three CO molecules. Magnetic moment, charge, and position of the substitutional atom are also strongly affected by the CO adsorption in most cases. Magnetic moment is stepwise reduced to zero, and the outward displacement of the substitutional atom increased.

\end{abstract}




The electrochemical reduction reaction of CO$_2$, referred to as CO2RR, is a promising way of converting CO$_2$ into useful products.\cite{Hori_2002,Hori_2003}
Depending on the electrode material, various products can be formed, {\it e.g.}, formic acid, CO, and further reduced products including methane and even  C$_{2}$ products, such as ethylene (see, for example, \citet{Rhimi_2024} for a recent review).
Remarkably, more ethylene than C$_1$ products are formed at the Cu(100) surface for a certain range of applied voltage.\cite{Schouten_2012}
A two-parameter descriptor based on the binding energy of a CO admolecule and of an H adatom on the surface of the catalyst has been proposed as a measure for predicting the products formed in CO2RR.\cite{Hussain_2018,Bagger_2019}

Copper electrodes are particularly promising, as the CO intermediate binds strongly enough to the surface to become reduced further, while the coverage of hydrogen is low enough for the hydrogen evolution reaction not to dominate when the magnitude of the applied voltage is in the right window.\cite{Hussain_2018} Formation of formate is a competing reaction path that terminates as the anion is repelled from the negatively charged electrode surface.\cite{Van_den_Bossche_2021} 
The efficiency and selectivity for specific products need to be improved, though, in order for the process to become economical.

There have been reports of experiments where improved efficiency and/or selectivity is obtained by
adding other elements as substitutional impurities, for example, 
cobalt\cite{Bernal_2018,Yan_2021} and nickel.\cite{Xu_2020}
Even with a low surface concentration of the added element, the reaction pathway and overpotential needed for CO2RR can be modified substantially.
Such copper-based surfaces with a small amount of substitutional impurities can be regarded as examples of a wider class of catalysts referred to as single-atom alloys (SAAs) where individual atoms of more reactive element, such as Pt, Fe, Co, or Ni, are embedded in a matrix of a less reactive element, copper in the present case.\cite{Hannagan_2020, Reocreux_2022,Schumann_2024}

In an effort to search for better CO2RR electrocatalysts, the adsorption energy of a CO molecule on the surface has been used as a key indicator, and the calculations are typically carried out using density functional theory (DFT) (see, for example, \citet{Liu_2023}). From a large number of such calculations, 
machine learned models are being developed for predicting the adsorption energy of a CO molecule on copper surfaces modified by introducing a wide range of other elements.\cite{Salomone_2023}
Following the Sabatier principle, a good catalyst should have the right balance between keeping the intermediates on the surface and releasing the products. 
The binding of a CO molecule is taken to be a good descriptor for any intermediate where a C atom binds to the surface, such as, for example, OCOH, COH, and CH$_3$, and
the search for a better CO2RR catalyst has therefore focused on finding the optimal value for the binding energy of a single CO molecule on the catalyst surface.

There are, however, some indications that more than one CO molecule can adsorb on a transition metal substitution atom on a copper surface.
Adsorption of two CO molecules has been experimentally observed for Cu surfaces doped with either Co \cite{Eren_2018} or with Rh,\cite{Wang_2022} and this has been supported by DFT calculations.
This indicates that in order to apply the Sabatier principle correctly and assess the catalytic performance, one needs to consider the adsorption of whatever number of CO molecules is expected to be present under reaction conditions, not just the adsorption of the first CO. It is then more likely that the binding energy of the last adsorbed CO molecule on the substitutional atom will be a useful descriptor of the CO2RR catalytic performance.

Here, the adsorption of multiple CO molecules on substitutional atoms from the first row of transition metals left of copper in the periodic table are studied using DFT calculations for Cu(100) and Cu(111) surfaces. 
The structure and energy for up to four CO admolecules are found, and in some cases the 4$^{th}$ CO admolecule binds more strongly on the substitutional atom than on the pure copper surface. 
This shows that it is essential to consider the binding of multiple CO molecules in order to assess CO2RR catalytic performance. 
Various properties such as magnetic moment, charge transfer and outward displacement of the substitutional atom are strongly affected by the CO adsorption.
In a broader context, our results can be relevant for SAA catalysts of other reactions in that one should be careful to consider the possibility of multiple admolecules binding to the same site.



Figure 1 shows the structure obtained when a vanadium atom is placed in the Cu(100) surface layer and up to 4 CO molecules bind to it. Even the fourth CO molecule binds more strongly to the V atom than to the clean Cu(100) surface. However, on the Cu(111) surface, the 4$^{th}$ CO molecule is bound more strongly on the copper surface than on the V
atom.

\begin{figure}
    \centering
    \includegraphics[width=1\linewidth]{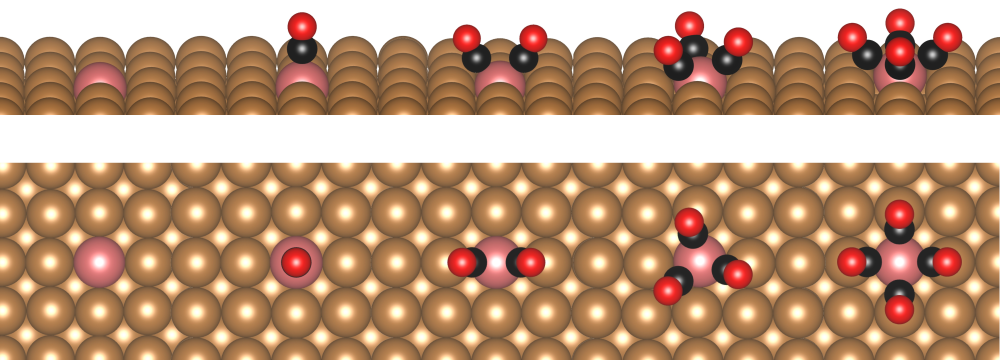}
    \caption{Minimum energy structure of the Cu(100) surface where one of the Cu atoms has been replaced with a V atom (pink) and then 1, 2, 3, and 4 CO molecules adsorbed on the V atom (O atom red, C atom black). Even the fourth CO molecule is bound more strongly to the V atom than on the pure Cu(100) surface.}
    \label{fig:surface}
\end{figure}

\begin{figure}[H]
    \centering
    \includegraphics[width=0.95\linewidth]{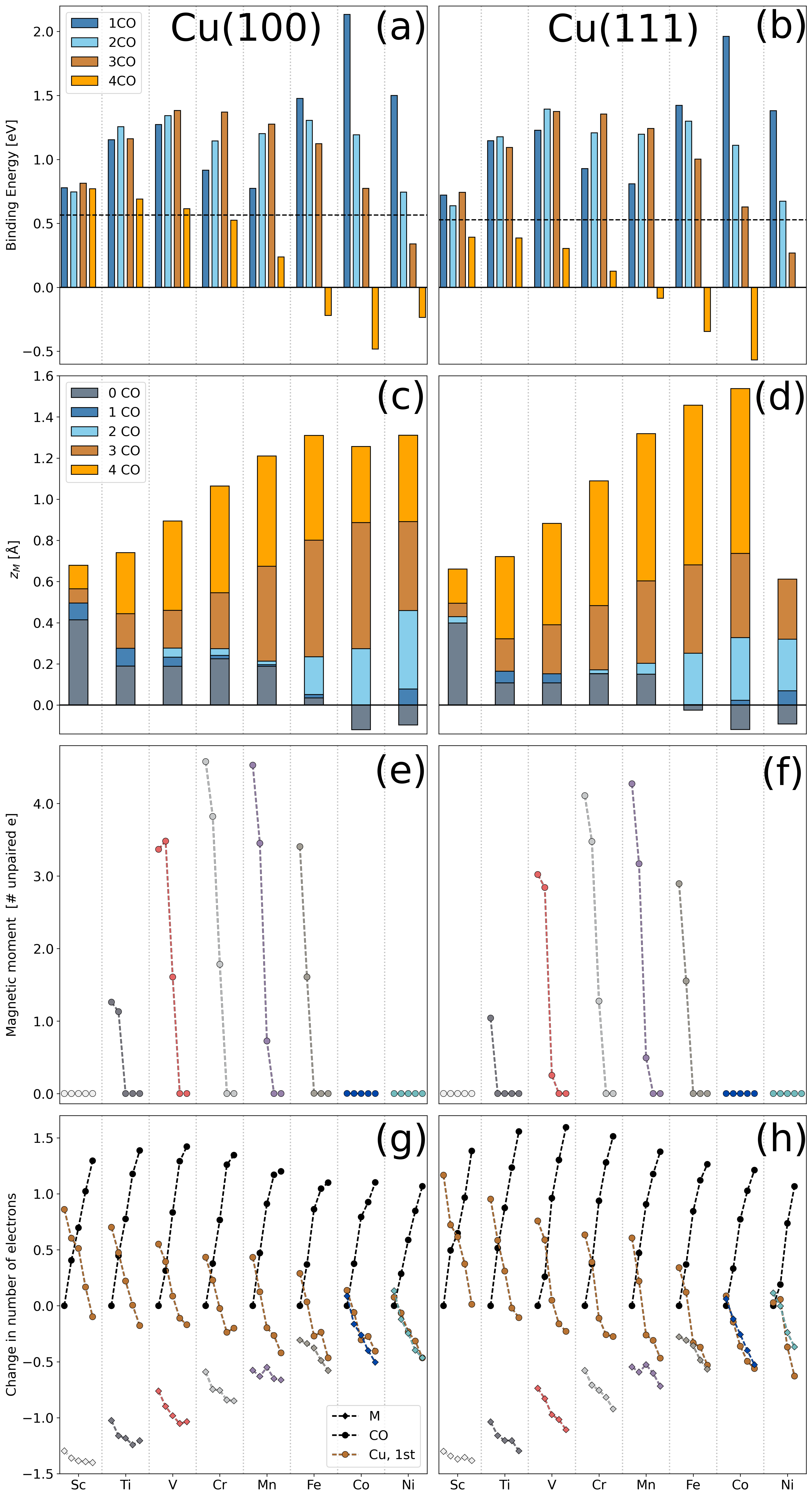}
    \caption{Left:  Cu(100). Right: Cu(111).
    (a and b): Differential binding energy, $E_\mathrm{b\text{-}n^{th}\,CO/M@Cu}$, of a CO molecule adsorbed on a substituent atom, M.
    Up to 4 CO molecules are adsorbed on a single M atom. The black dashed lines indicate the binding energy of a CO molecule on the pure copper surfaces. 
    (c and d): Position of the M atom with respect to the surface plane (z-coordinate) when no and up to 4 CO molecules are adsorbed.
    (e and f): Total magnetic moment of the simulated system, (CO)$_{n}$-M@Cu, indicating the number of unpaired electrons. 
    (g and h): Change in the number of electrons associated with the M atom, the first neighbouring surface Cu atoms, and CO molecules adsorbed on M for varying number of adsorbed CO molecules, based on Bader analysis.
    }
    \label{fig:main}
\end{figure}

Figure \ref{fig:main}(a) shows the calculated 
differential binding energy, $E_\mathrm{b\text{-}n^{th}\,CO/M@Cu}$ (see Methods), for CO on Sc, Ti, V, Cr, Mn, Fe, Co, and Ni substitutional atom in Cu(100). Analogous results for Cu(111) are shown in Figure \ref{fig:main}(b).
The binding energy of a single CO molecule on the corresponding pure Cu surface is shown for comparison. 
The differential binding energy 
gives the change in the total energy 
as the n$^{th}$ CO molecule coming from the gas phase adsorbs on the substitutional atom with n-1 CO molecules present already.
The differential binding energy in most cases depends strongly on the number of CO admolecules.
%
For Fe, Co, and Ni, the differential binding energy decreases as more CO molecules are adsorbed. This is the expected trend; the more bonds are formed to a given atom, the weaker each bond is. For both the (111) and (100) facets, the Fe and Co atoms can bind three CO molecules more strongly than the pure copper surface. 
While there is a local minimum 
on the energy surface
for the adsorption of a 4$^{th}$ molecule, the energy of the system is lowered by placing the molecule on the copper surface far from the substitutional atom.
For Ni, 2 CO molecules adsorb more strongly than on the pure copper surface.
The variation in differential binding energy is particularly large for Co and Ni, where the second CO molecule has a binding energy nearly 1 eV lower binding energy than the first one.

An opposite trend is found for the differential binding energy on Cr and Mn atoms, as it increases for the first three
CO molecules. This is unusual. The 4$^{th}$ CO binds less strongly than on the pure copper surface. An analysis of the electronic structure changes associated with the CO adsorption is discussed below.
For Sc, Ti, and V the differential binding energy is rather constant for the first three CO admolecules, but decreases for the 4$^{th}$ one. Sc on the Cu(100) surface is an exception in that all four CO molecules have similar binding energy. 
The Sc, Ti and V atoms in a Cu(100) surface are special in that the 4$^{th}$ CO admolecule binds more strongly than on the pure copper surface.

The substitutional atoms are displaced outward with respect to the surface plane as the CO molecules adsorb. This is shown in
Figures \ref{fig:main}(c) and (d) for the Cu(100) and Cu(111) surfaces, respectively. The position of the substitutional atom in the direction normal to the surface
with respect to the average position of the Cu atoms in the surface layer of the M@Cu system is displayed.
Sc being the largest of the first row transition metal atoms, it is displaced outwards by 0.4 {\AA} even in the absence of CO admolecules. The adsorption of CO increases the displacement only slightly, by {\it ca.} 0.1 {\AA}, except for the second CO admolecule.
On the other end of the sequence, the Co and Ni atoms are relatively smaller and they are {\it ca.} 0.1 {\AA} lower than the surface Cu atoms in the absence of CO. In all cases, the adsorption of CO leads to outward displacement of the substitutional atoms. As the 4$^{th}$ CO molecule is adsorbed, the outward displacement can be greater than 1.5 {\AA}.
We note that it is in some cases important to first displace the substitutional atom outwards as a new CO molecule is added before carrying out local energy minimization in order to find the global or local energy minimum corresponding to the increased number of CO admolecules.

The calculated magnetic moment, indicating the number of unpaired d-electrons, is shown in 
figures \ref{fig:main}(e) and (f). As the number of CO admolecules increases, the magnetic moment in most cases decreases strongly. 
The reason for this appears to be a shift in the energy of the d-orbitals of the substitutional atoms as the CO molecules adsorb. 
This is to be expected since CO is a strong ligand.
%
For Sc, Co and Ni the magnetic moment is zero and remains unchanged.
The transfer of electrons is estimated using Bader analysis and the results shown in
figures \ref{fig:main}(g) and (f). 
Electrons get transferred to the CO admolecules, mainly from the adjacent Cu atoms but also from the substitutional atom.
This electron transfer likely contributes also to the changes in the magnetic moment.



The calculated results presented here 
show that several CO molecules can be bound to a first row transition metal substituent atom in (111) and (100) copper surfaces.
The differential binding energy varies in most cases significantly as more CO molecules are adsorbed. 
This means that the binding of a single CO molecule is not likely to be a good descriptor for CO2RR activity. 
Copper is known to be a good electrocatalyst for CO$_2$ reduction because the affinity for admolecules is just about right, strong enough for the CO to remain on the surface at room temperature  while the binding of H-adatoms is weak enough for the hydrogen evolution reaction not to dominate.\cite{Hussain_2018} 
The C-C bond formation is known to be particularly efficient on Cu(100), and an interesting question is the possible reaction of a CO bound to a substitutional atom with a CO molecule bound on nearby sites formed by Cu atoms. 
By analogy, only the rather weakly bound CO molecules on the substitutional atoms are most likely to react to form C$_2$ products. The strongly bound CO molecules are unlikely to be reactive.

For the Cu(100) surface, it is the 4$^{th}$ CO molecule on Ti, V, or Cr that has differential binding energy close to that of CO on the pure Cu(100).
The 3$^{rd}$ CO molecule on Co and the 2$^{nd}$ on Ni are close but bind a bit too strongly, as are all CO molecules adsorbed on Sc. 
None of the CO molecules adsorbed on a Mn or Fe substitutional atom have differential binding energy in the right range.
The first CO bound to Mn is only a bit too strongly bound, but the second and third CO molecules bind so strongly that the most stable situation is three strongly bound and chemically inactive CO admolecules.
For the Cu(111) surface, only the 3$^{rd}$ CO molecule on Co and the 2$^{nd}$ on Ni have differential binding energy close to that of the pure copper surface. 
A stable structure with four CO molecules bound to the Ni atom in Cu(111) was not found.
%

The anomalous trend of increasing differential binding energy as additional CO molecules are adsorbed on Ti, V, Cr, and Mn substitutional atoms seems to be associated with changes in the magnetic moments; namely a reduction in the number of unpaired electrons. The non-magnetic Sc, Co and Ni substitutional atoms in the copper surfaces show the expected trend of decreasing differential binding energy as the number of CO molecules is increased. A more detailed analysis of the bonding and electronic structure is called for to obtain a better understanding of these results.

The large outward displacement of the substitutional atoms as 3 and even 4 CO molecules are adsorbed, on the order of an {\AA}ngström, raises the question of whether the substitutional atom may leave the surface altogether as a carbonyl complex. This is an important issue, as the stability of electrocatalysts is equally important as efficiency. Again, further studies of this are in order and will be the subject of later reports.

The sites on the copper surface in the vicinity of the substitutional atom can also be modified as compared with analogous sites on a pure surface. This in itself could lead to changes in CO2RR activity even if none of the CO molecules adsorbed on the substitutional atom are active. This requires additional calculations and will be addressed in later studies.



The main outcome of the studies presented here is a warning that should be kept in mind in the large computational effort now being undertaken in a search for an improved CO2RR catalyst by adding a minority component to copper and in studies of SAA catalysts in general.
The binding energy of a single CO molecule on a transition metal substitutional atom on a copper surface is not likely to be a good descriptor for CO2RR activity. Multiple CO molecules can be bound, in most cases too strongly to be reactive. The 3$^{rd}$ or even the 4$^{th}$ CO molecule adsorbed on the substitutional atom is more likely to be reactive than the first one. In some cases the differential binding energy increases as more CO molecules are adsorbed, unlike the usual trend of bonds weakening as more bonds are formed. The reason for this appears to be related to the magnetic properties, i.e., reduction in the number of unpaired electrons. 


\section{Methods}

The calculations were carried out using spin-polarized density functional theory (DFT).\cite{Kohn_1999} 
A plane-wave basis set with 450 eV cutoff energy was used to represent the valence electrons, in a gamma-centered k-point grid of $4\times4\times1$, while the effect of the  
inner electrons was included using the projector augmented wave (PAW) approach.\cite{Blochl_1994} 
The revised Perdew-Burke-Ernzerhof (RPBE) functional approximation\cite{Hammer_1999} 
was used as it is parametrized to give quite good agreement between calculated and measured energy of molecular adsorption on solid surfaces. 
The binding energy of a CO molecule on the clean Cu(100) and Cu(111) surfaces was found to be 0.57 eV and 0.53 eV, as compared with the experimental values of 0.53 eV and 0.49 eV.\cite{Vollmer_2001} 
The calculated lattice constant of copper crystal comes out to be 3.70\,Å, 2\% larger than the experimental value.

The minimization of the energy is carried out using a conjugate gradient optimizer until the magnitude of all atomic forces drops below  0.01 eV/Å. The electronic structure optimization is carried out until self-consistency is reached at a level of $10^{-6}$ eV. 
The calculations are carried out using the Vienna {\it ab initio} simulation package (VASP).\cite{VASP_3,VASP_4}

The substitutional atom is termed M and the copper surface with a substitutional atom is referred to as M@Cu.
In order to find the energy minimum for more than one CO molecule adsorbed on the substitutional atom, it sometimes needs to be lifted higher above the surface plane before local minimization of the energy is applied with respect to atom coordinates. A comparison is then made with the energy of the system when the added CO molecule is placed on a pure copper surface to determine whether adsorption on the substitutional atom is preferable.

The differential binding energy, i.e. energy gain/cost of adding the n$^{th}$ CO molecule coming from the gas phase, is calculated as
\begin{equation*}
    E_\mathrm{b\text{-}n^{th}\,CO/M@Cu}  = E_{\mathrm{(CO)_{n}/M@Cu}} - E_{\mathrm{(CO)_{n\text{-}1}/M@Cu}} - E_\mathrm{CO},
\end{equation*}
where $E_{\mathrm{CO}}$ is the energy of an isolated CO molecule, and $E_{\mathrm{(CO)_{n}/M@Cu}}$ 
is the values obtained for the total energy of the M@Cu surface with n 
CO molecules adsorbed. 
The energy of the converged $\mathrm{(CO)_n/M@Cu}$ structures depends on the configuration of the CO molecules, with some surfaces having energy minimas for the CO molecules either pointing toward the hollow sites, bridge sites, or pointing away perpendicular from the surface. All these configurations were considered with the initial structures of the structural optimization calculations.

The charge on the atoms was estimated using the Bader definition\cite{Bader_1990} implemented using a grid approach.\cite{Henkelman_2006}




\begin{acknowledgement}

This work was funded by the Icelandic Research Fund (Grant No. 207283-053) and the EU’s Horizon 2021 programme under the Marie Skłodowska-Curie Doctoral Networks (MSCA-DN) Grant Agreement No. 101072830 (ECOMATES).
We thank Giancarlo Cicero,
Kuber Rawat, and Kun Zhao for helpful discussions.

Views and opinions expressed are those of the authors and do not necessarily reflect the official policy or position of the European Union
or the Granting Authority, and neither can be held responsible for them.

\end{acknowledgement}




\bibliography{references_abbr}

\end{document}


\noindent
{\bf Note on the geometry of an adsorbed CO}

The first CO molecule adsorbed on V, Fe, Co and Ni substitutional atom is directed along the surface normal, but on Sc, Ti, Cr and Mn a tilted configuration has lower energy. The energy difference between tilt and normal direction is on the order of 0.2 eV or less. For Sc, Ti, Cr the first adsorbed CO tilts towards the bridge site on the Cu(100) surface, and the HCP site on the Cu(111) surface. For Mn, the first adsorbed CO tilts towards the hollow site of the Cu(100) surface, and the HCP site on the Cu(111) surface.

\begin{figure*}[]
    \centering
    \begin{subfigure}[b]{0.5\textwidth}
        \centering
        \includegraphics[width=0.9\linewidth]{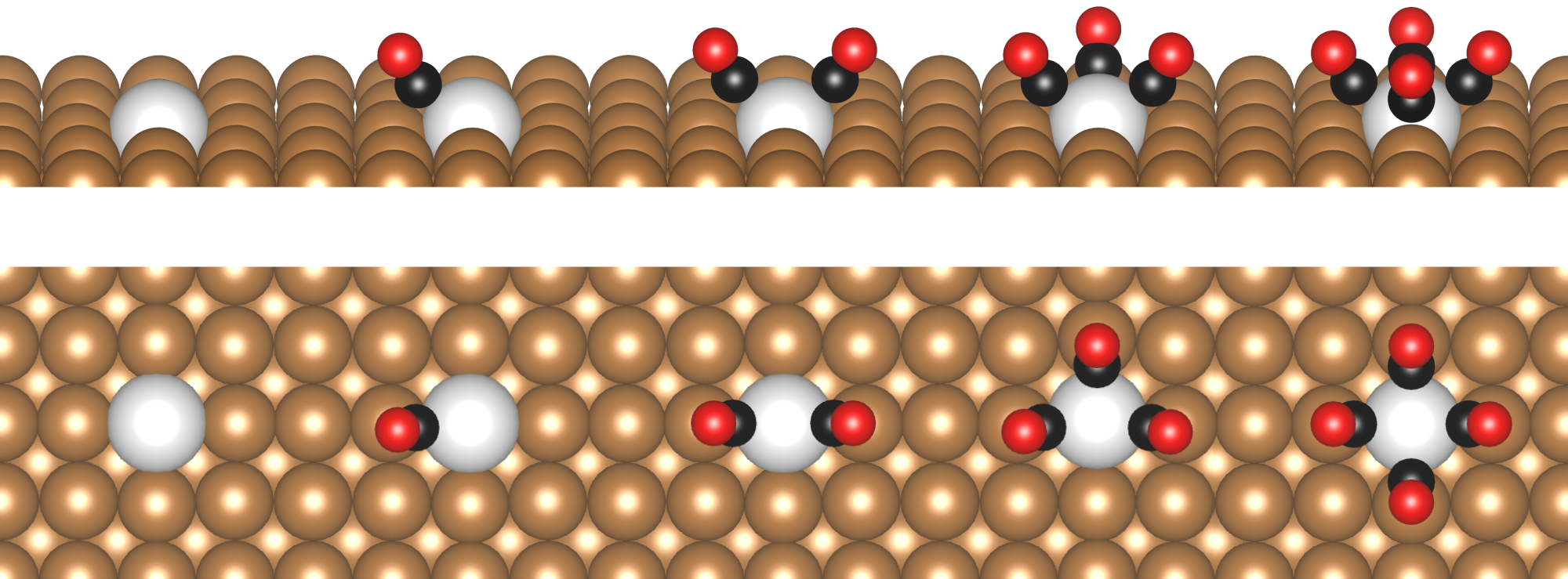}
        \caption{Sc@Cu(100)}
    \end{subfigure}%
    \begin{subfigure}[b]{0.5\textwidth}
        \centering
        \includegraphics[width=0.9\linewidth]{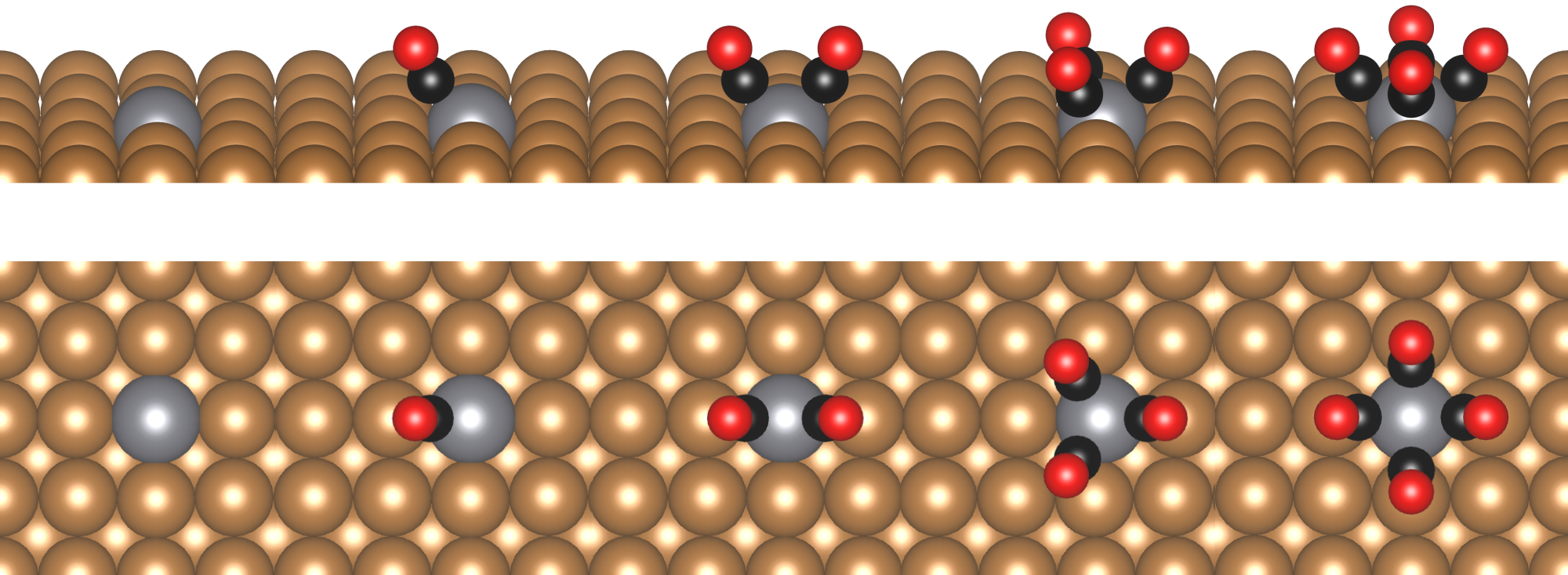}
        \caption{Ti@Cu(100)}
    \end{subfigure}
    \begin{subfigure}[b]{0.5\textwidth}
        \centering
        \includegraphics[width=0.9\linewidth]{figures/figure1.png}
        \caption{V@Cu(100)}
    \end{subfigure}%
    \begin{subfigure}[b]{0.5\textwidth}
        \centering
        \includegraphics[width=0.9\linewidth]{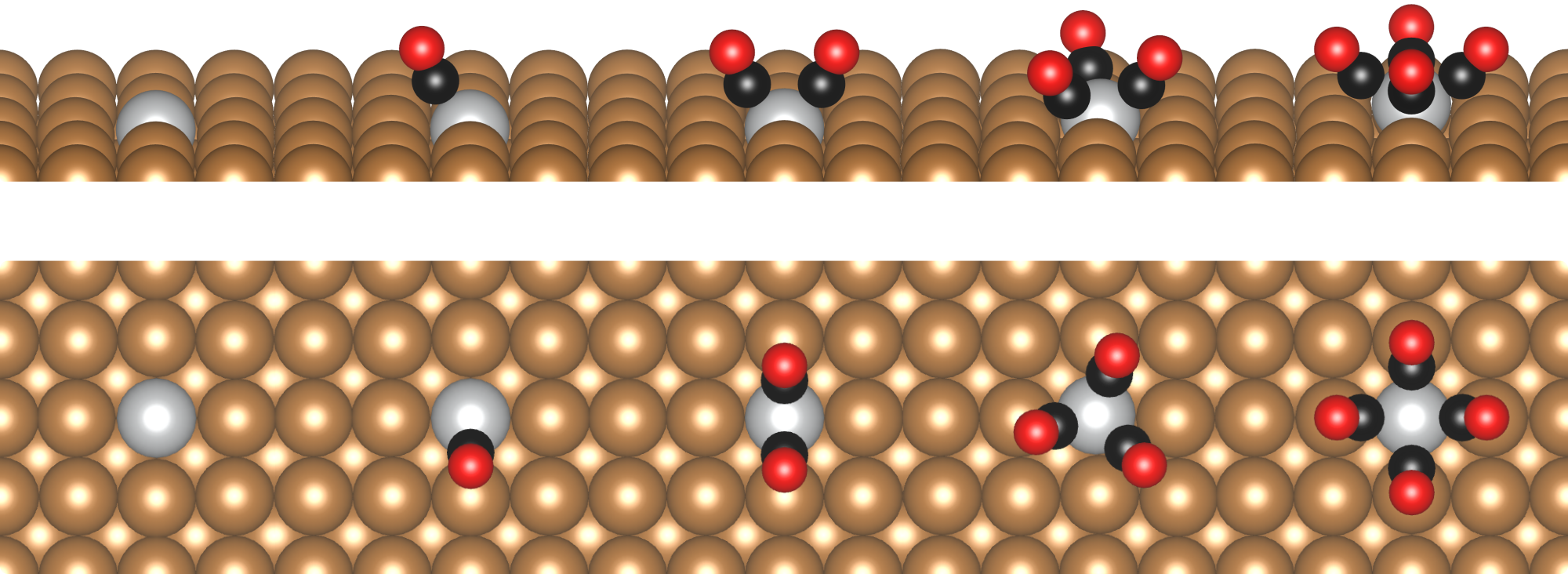}
        \caption{Cr@Cu(100)}
    \end{subfigure}
    \begin{subfigure}[b]{0.5\textwidth}
        \centering
        \includegraphics[width=0.9\linewidth]{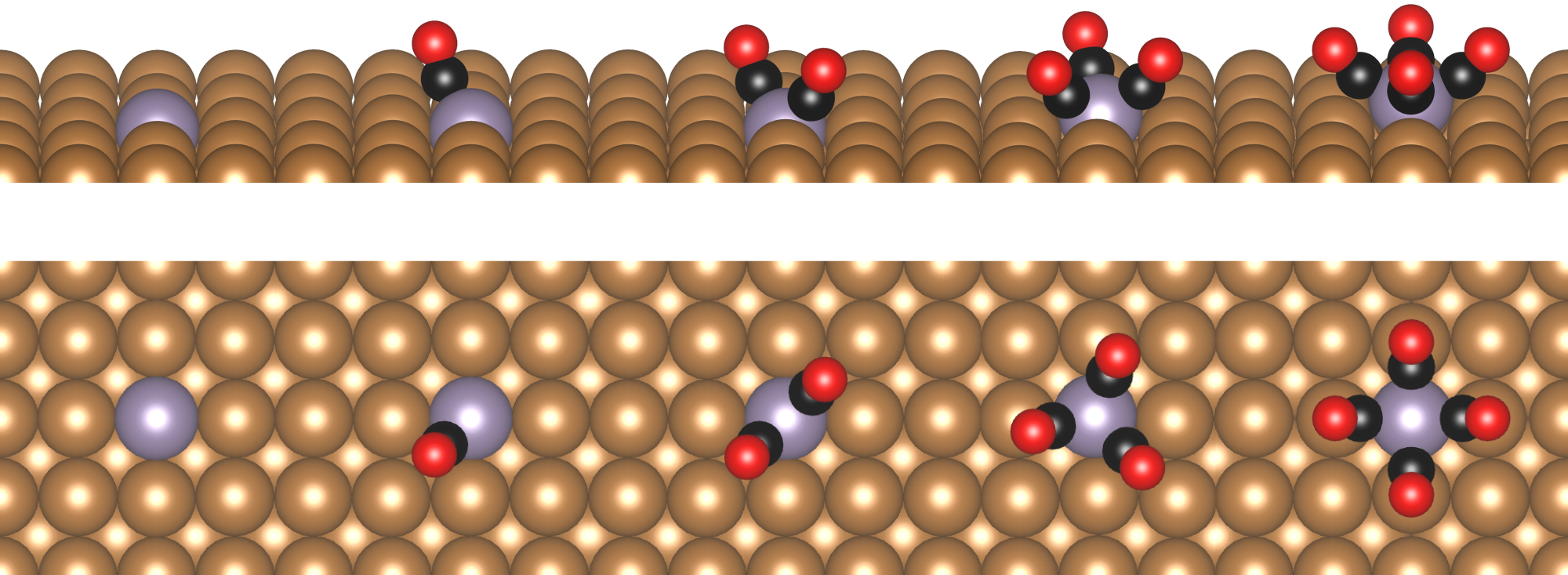}
        \caption{Mn@Cu(100)}
    \end{subfigure}%
    \begin{subfigure}[b]{0.5\textwidth}
        \centering
        \includegraphics[width=0.9\linewidth]{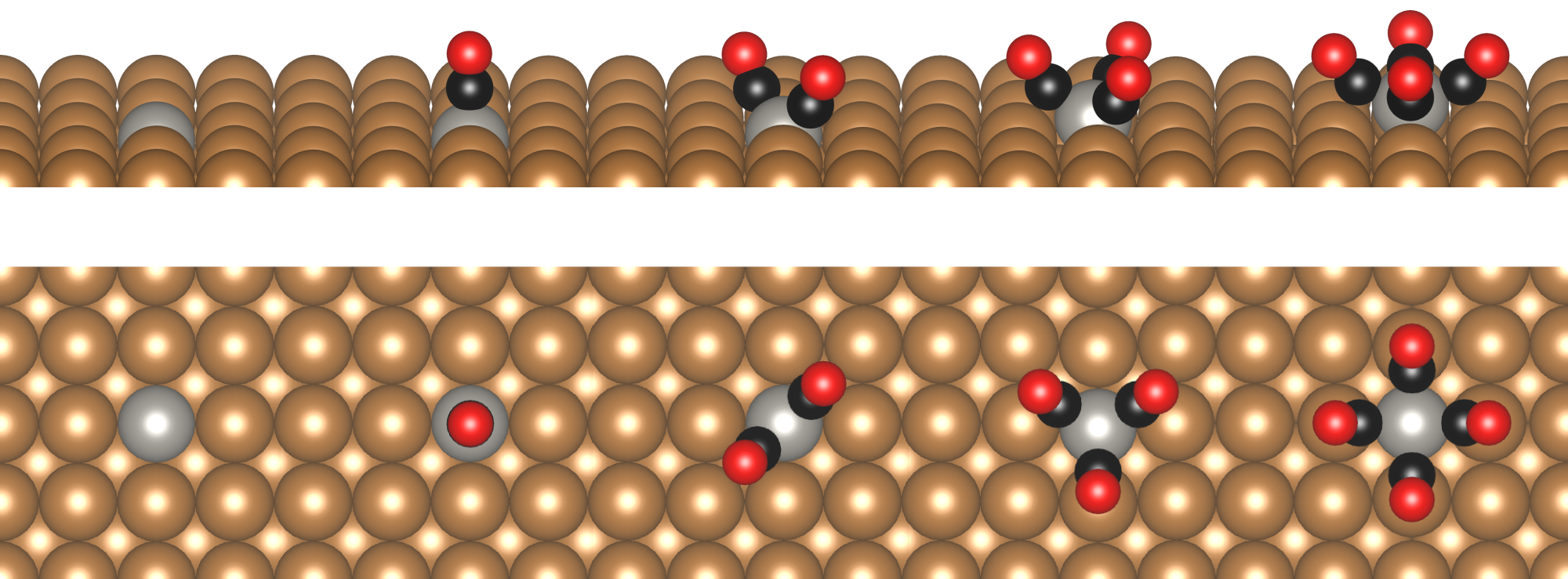}
        \caption{Fe@Cu(100)}
    \end{subfigure}
    \begin{subfigure}[b]{0.5\textwidth}
        \centering
        \includegraphics[width=0.9\linewidth]{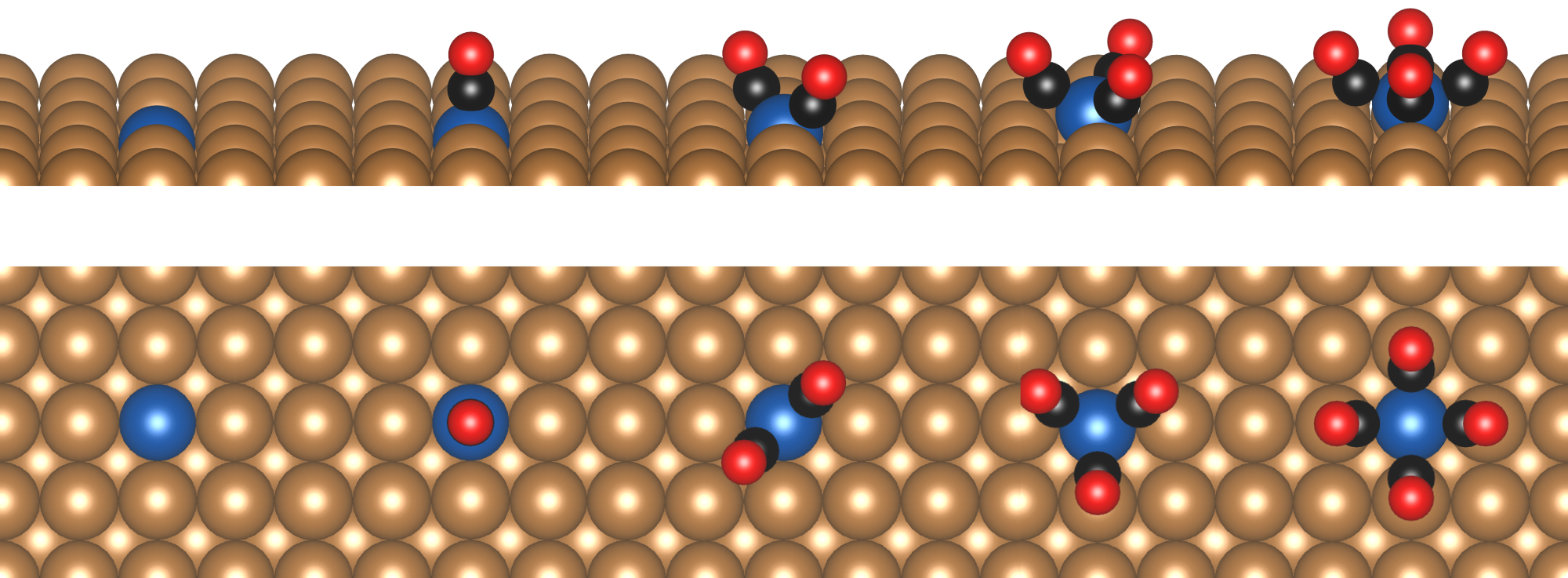}
        \caption{Co@Cu(100)}
    \end{subfigure}%
    \begin{subfigure}[b]{0.5\textwidth}
        \centering
        \includegraphics[width=0.9\linewidth]{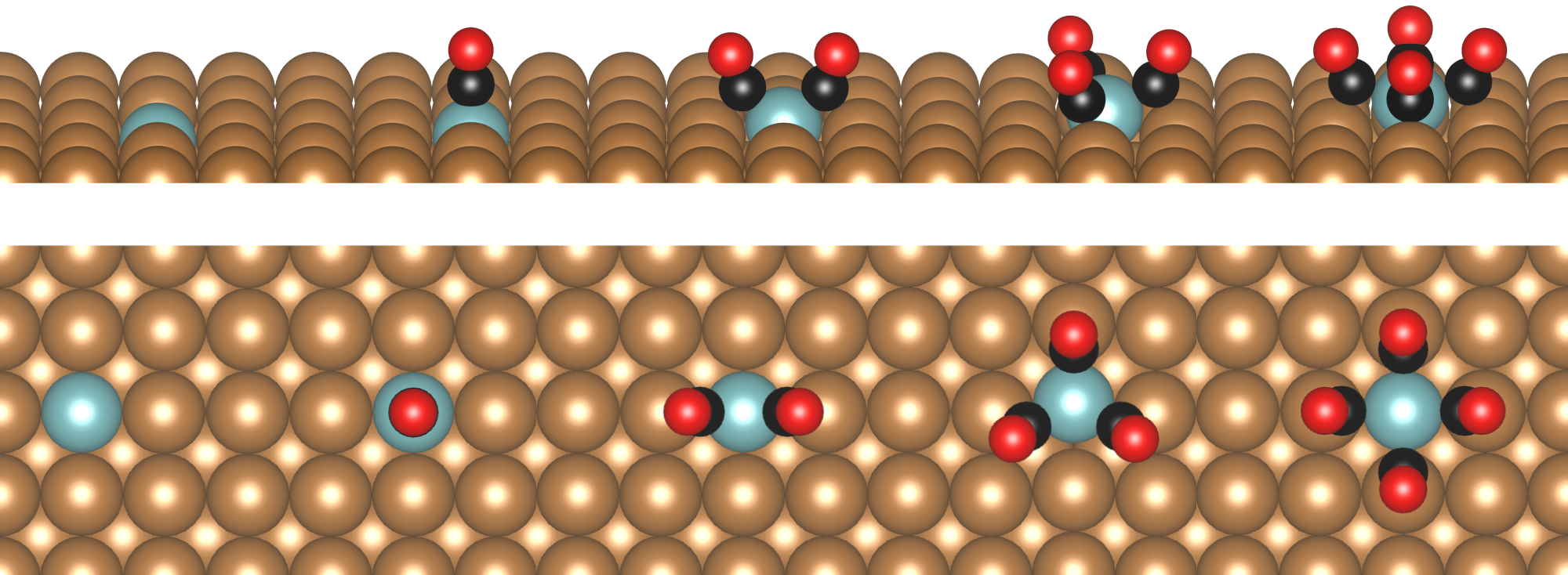}
        \caption{Ni@Cu(100)}
    \end{subfigure}
    \begin{subfigure}[b]{0.5\textwidth}
        \centering
        \includegraphics[width=0.7\linewidth]{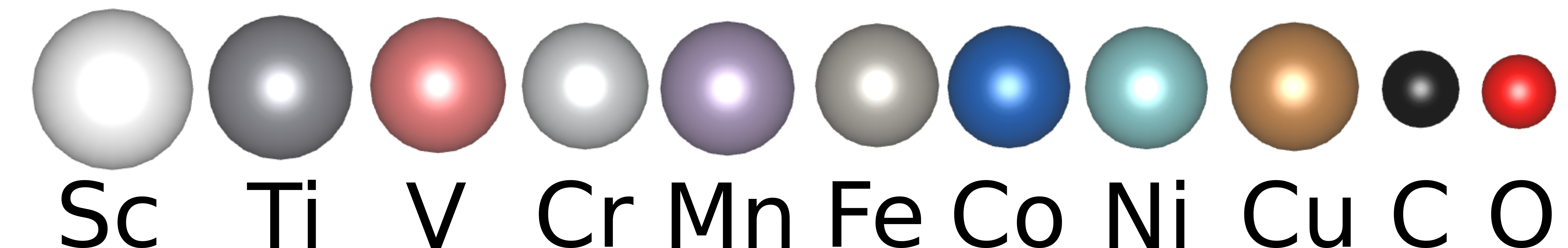}
    \end{subfigure}
    \caption{Minimum energy structures of (a) Sc, (b) Ti, (c) V, (d) Cr, (e) Mn, (f) Fe, (g) Co and (h) Ni on the Cu(100) surfaces with 0-4 CO molecules adsorbed to the substituent atom}
    \label{fig:M@Cu(100)}
\end{figure*}

\begin{figure*}[]
    \centering
    \begin{subfigure}[b]{0.5\textwidth}
        \centering
        \includegraphics[width=0.9\linewidth]{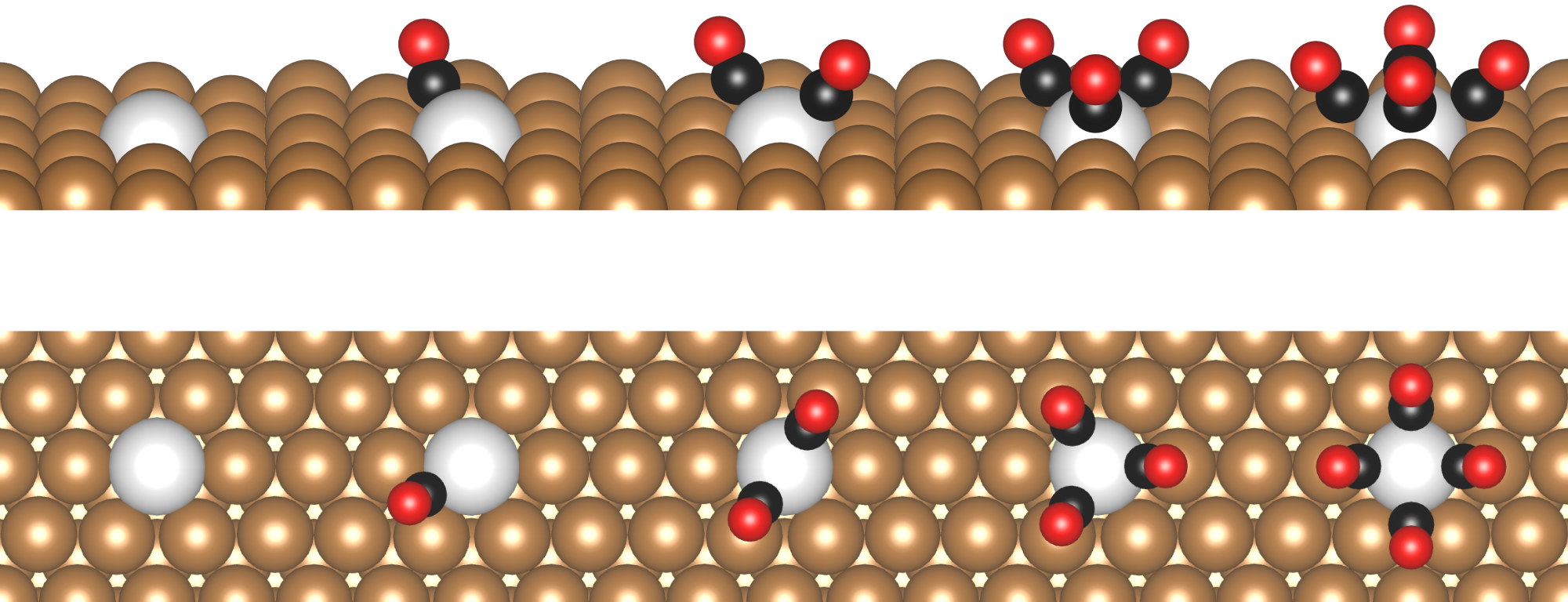}
        \caption{Sc@Cu(111)}
    \end{subfigure}%
    \begin{subfigure}[b]{0.5\textwidth}
        \centering
        \includegraphics[width=0.9\linewidth]{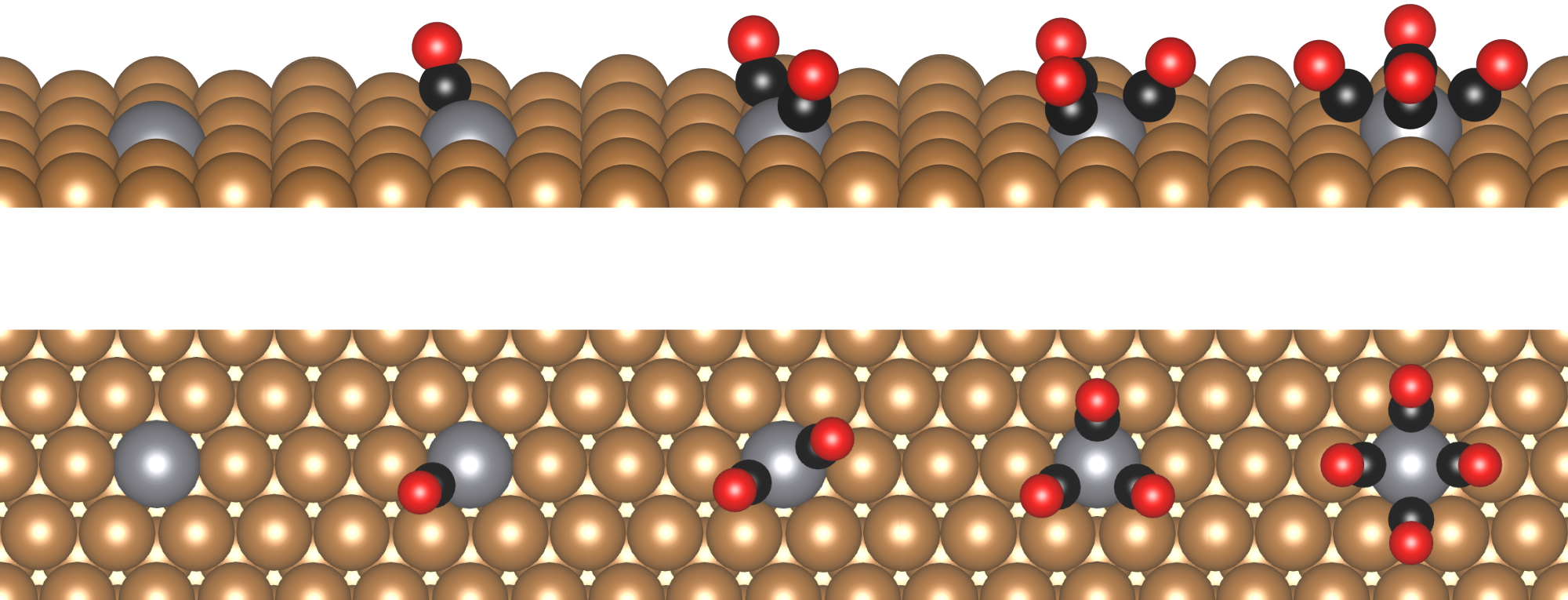}
        \caption{Ti@Cu(111)}
    \end{subfigure}
    \begin{subfigure}[b]{0.5\textwidth}
        \centering
        \includegraphics[width=0.9\linewidth]{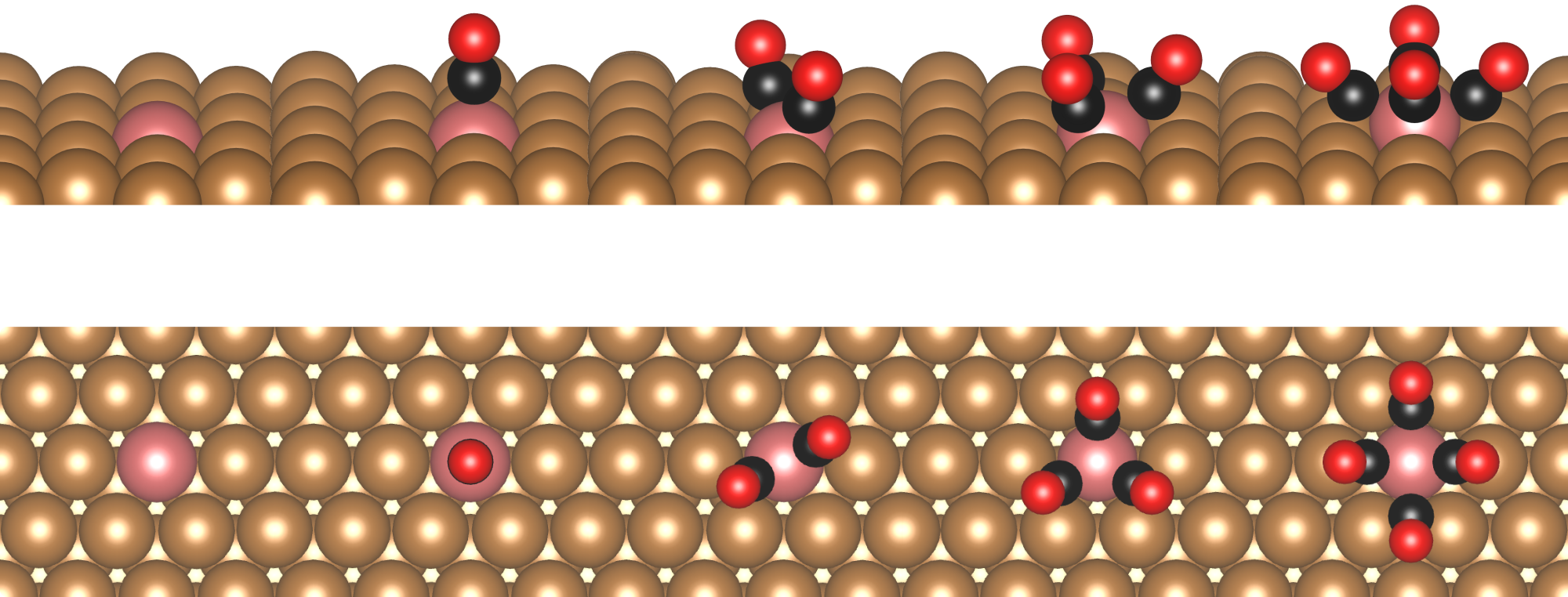}
        \caption{V@Cu(111)}
    \end{subfigure}%
    \begin{subfigure}[b]{0.5\textwidth}
        \centering
        \includegraphics[width=0.9\linewidth]{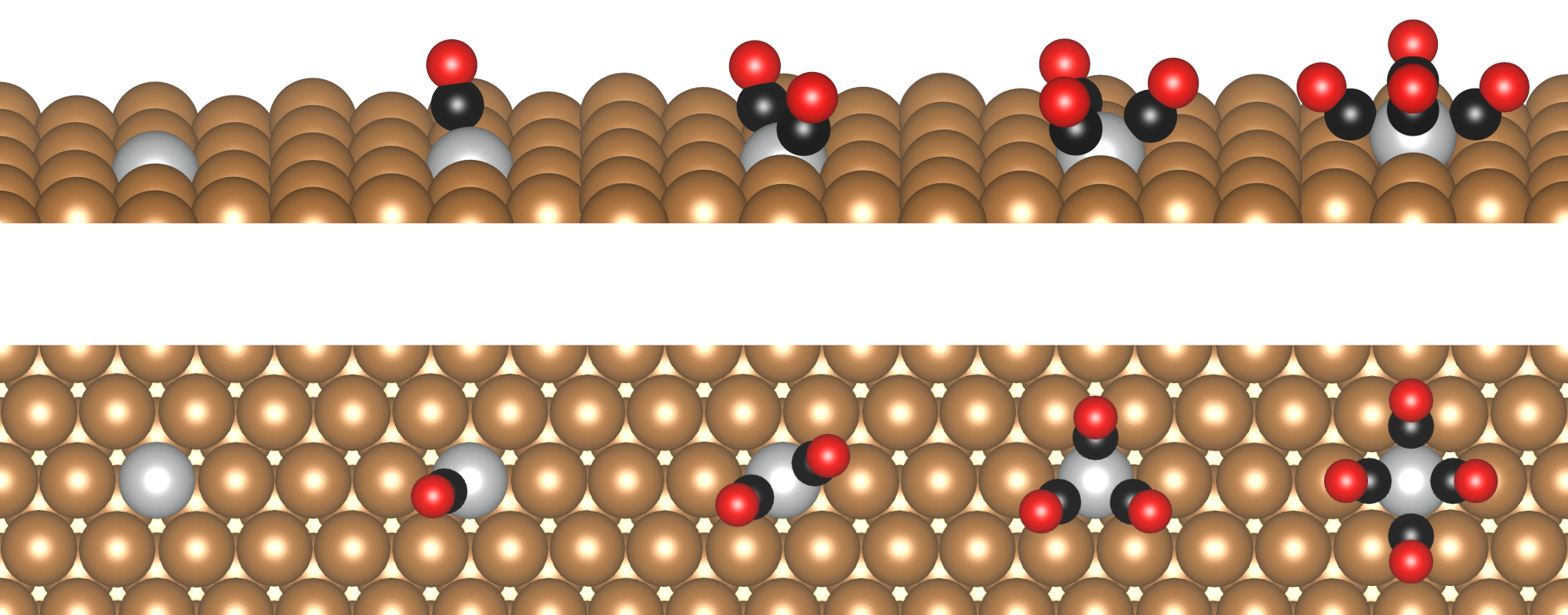}
        \caption{Cr@Cu(111)}
    \end{subfigure}
    \begin{subfigure}[b]{0.5\textwidth}
        \centering
        \includegraphics[width=0.9\linewidth]{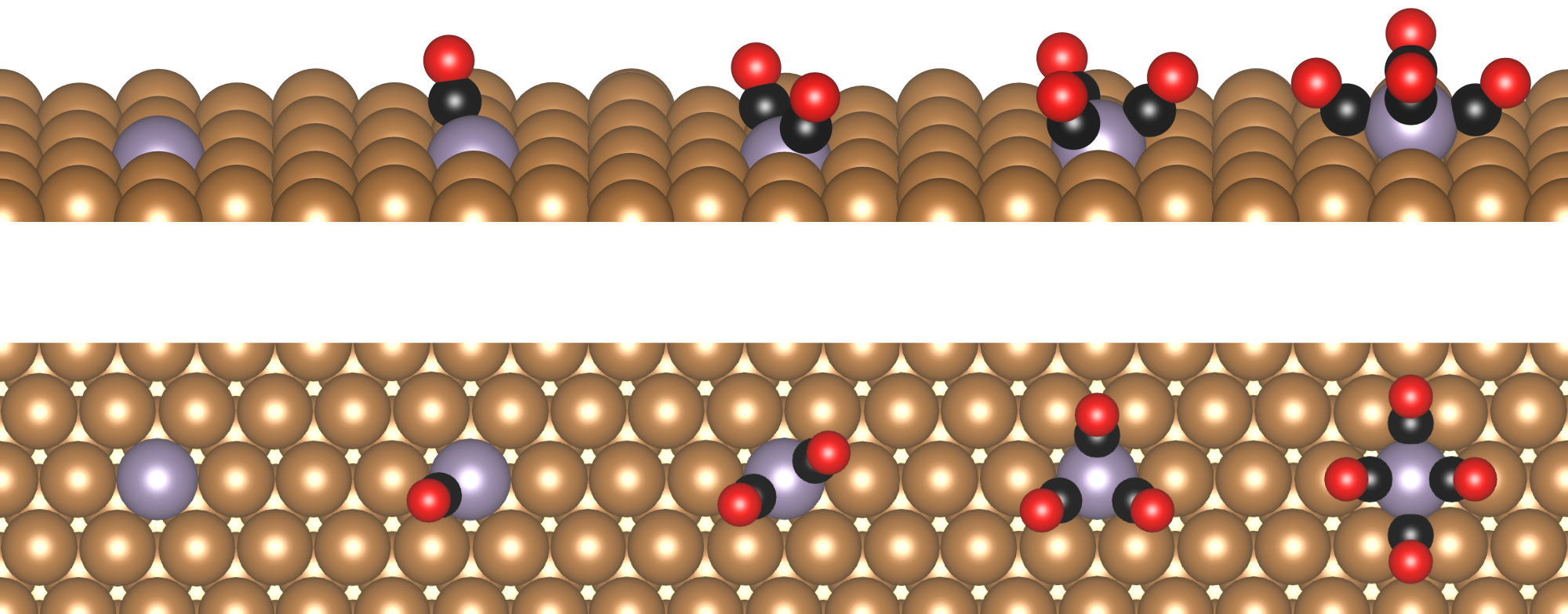}
        \caption{Mn@Cu(111)}
    \end{subfigure}%
    \begin{subfigure}[b]{0.5\textwidth}
        \centering
        \includegraphics[width=0.9\linewidth]{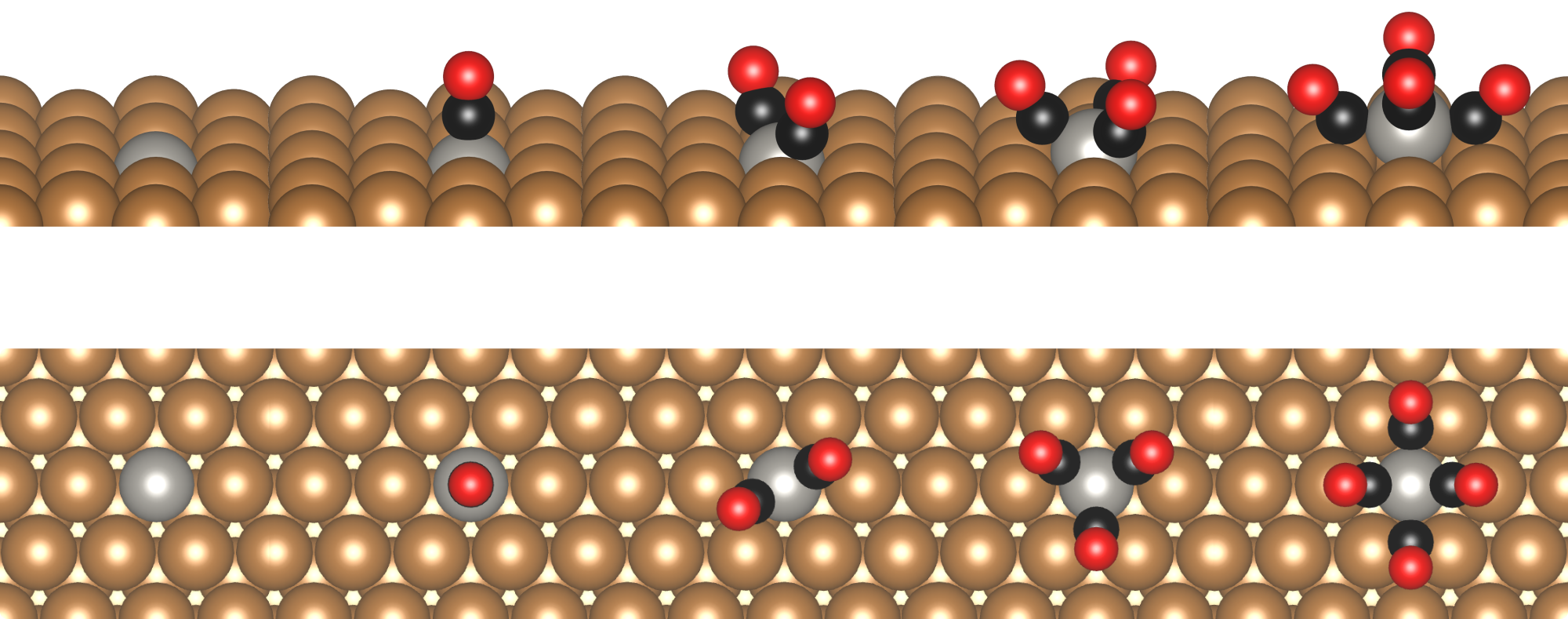}
        \caption{Fe@Cu(111)}
    \end{subfigure}
    \begin{subfigure}[b]{0.5\textwidth}
        \centering
        \includegraphics[width=0.9\linewidth]{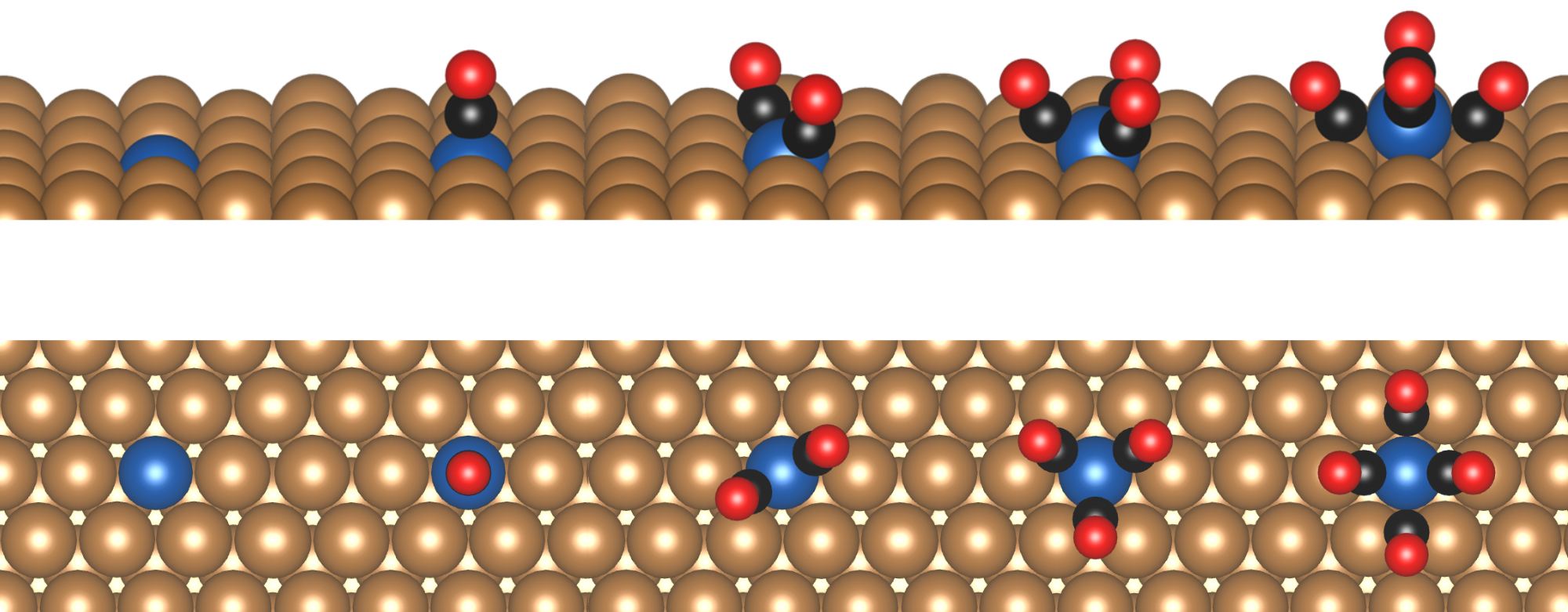}
        \caption{Co@Cu(111)}
    \end{subfigure}%
    \begin{subfigure}[b]{0.5\textwidth}
        \centering
        \includegraphics[width=0.73\linewidth]{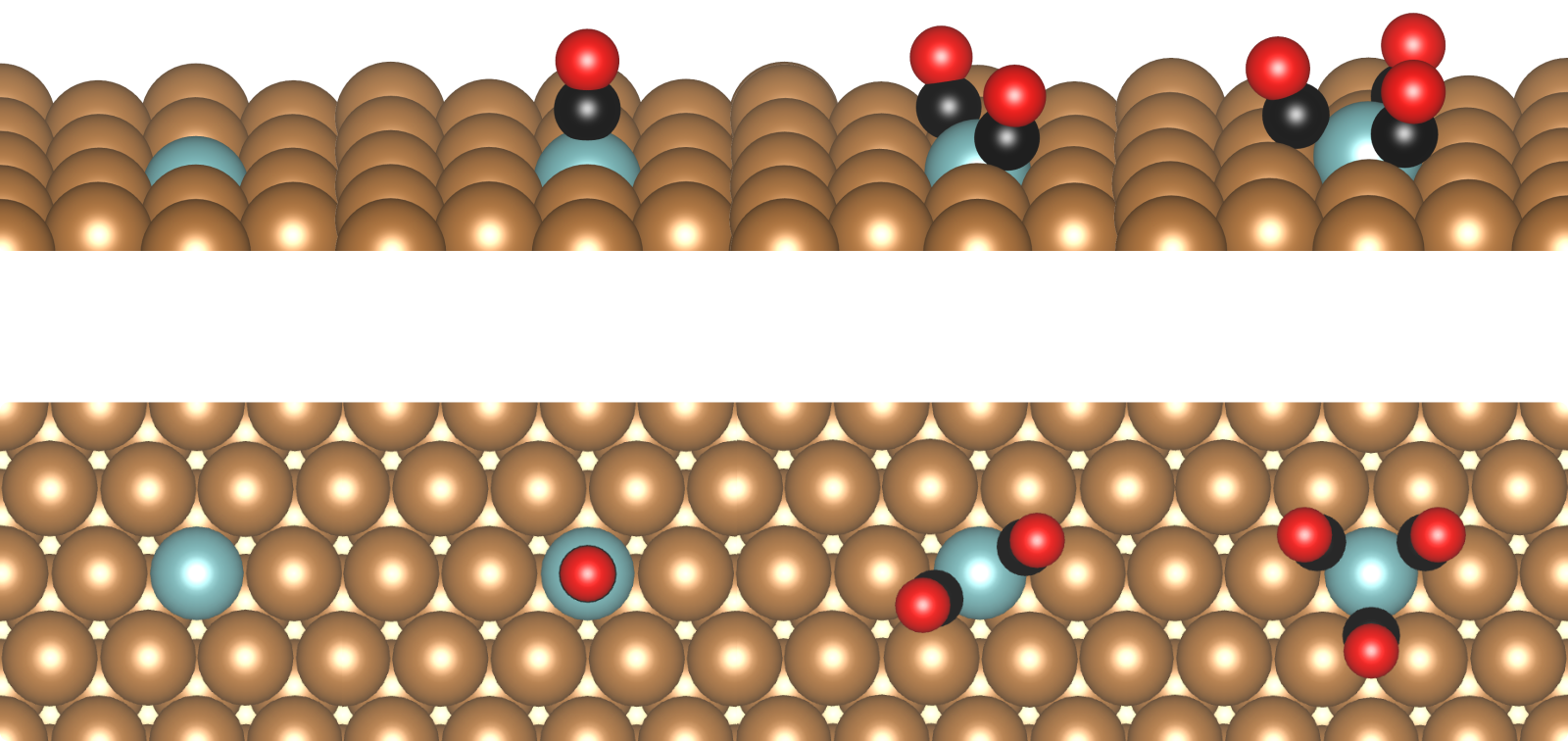}
        \caption{Ni@Cu(111)}
    \end{subfigure}
    \begin{subfigure}[b]{0.5\textwidth}
        \centering
        \includegraphics[width=0.7\linewidth]{figures/legend.png}
    \end{subfigure}
    \caption{Minimum energy structures of (a) Sc, (b) Ti, (c) V, (d) Cr, (e) Mn, (f) Fe, (g) Co and (h) Ni on the Cu(111) surfaces with 0-4 CO molecules adsorbed to the substituent atom}
    \label{fig:M@Cu(111)}
\end{figure*}

\begin{table}[]
    \centering
    \begin{tabular}{|c|c|c|c|c|c|c|c|c|} \hline
        M & \#CO & $E_\mathrm{b\text{-}n^{th}\,CO/M@Cu(100)}$ & $z_M$ & magmom & $\Delta q_M$ & $\Delta q_\mathrm{Cu, 1st}$ & $\Delta q_\mathrm{CO}$ \\ \hline
        \multirow{5}{*}{Sc} & 0 & -- & 0.414 \AA & 0.000 & 1.297 e & -0.861 e & -- \\ 
           & 1 & 0.779 eV & 0.496 \AA & 0.000 & 1.361 e & -0.604 e & -0.406 e \\ 
           & 2 & 0.747 eV & 0.486 \AA & 0.000 & 1.386 e & -0.513 e & -0.697 e \\ 
           & 3 & 0.814 eV & 0.565 \AA & 0.000 & 1.393 e & -0.167 e & -1.023 e \\ 
           & 4 & 0.770 eV & 0.679 \AA & 0.000 & 1.402 e &  0.098 e & -1.296 e \\ \hline
           
        \multirow{5}{*}{Ti} & 0 & -- & 0.190 \AA & 1.261 & 1.025 e & -0.702 e & -- \\
           & 1 & 1.154 eV & 0.276 \AA & 1.130 & 1.162 e & -0.474 e & -0.448 e \\
           & 2 & 1.255 eV & 0.272 \AA & 0.000 & 1.184 e & -0.221 e & -0.776 e \\
           & 3 & 1.163 eV & 0.444 \AA & 0.000 & 1.242 e & -0.005 e & -1.178 e \\
           & 4 & 0.691 eV & 0.740 \AA & 0.000 & 1.206 e &  0.177 e & -1.387 e \\ \hline
           
        \multirow{5}{*}{V} & 0 & -- & 0.189 \AA & 3.368 & 0.762 e & -0.551 e & -- \\
          & 1 & 1.273 eV & 0.233 \AA & 3.482 & 0.899 e & -0.394 e & -0.314 e  \\
          & 2 & 1.347 eV & 0.285 \AA & 1.608 & 0.957 e & -0.085 e & -0.805 e \\
          & 3 & 1.378 eV & 0.460 \AA & 0.000 & 1.052 e &  0.111 e & -1.292 e \\
          & 4 & 0.615 eV & 0.894 \AA & 0.000 & 1.036 e &  0.169 e & -1.422 e \\ \hline
          
        \multirow{5}{*}{Cr} & 0 & -- & 0.225 \AA & 4.579 & 0.591 e & -0.434 e & -- \\
           & 1 & 0.916 eV & 0.241 \AA & 3.822 & 0.746 e & -0.229 e & -0.377 e \\
           & 2 & 1.144 eV & 0.274 \AA & 1.783 & 0.756 e &  0.024 e & -0.766 e \\
           & 3 & 1.370 eV & 0.545 \AA & 0.000 & 0.842 e &  0.237 e & -1.261 e \\
           & 4 & 0.525 eV & 1.064 \AA & 0.000 & 0.850 e &  0.200 e & -1.346 e \\ \hline
           
        \multirow{5}{*}{Mn} & 0 & -- & 0.188 \AA & 4.530 & 0.576 e & -0.433 e & -- \\
           & 1 & 0.773 eV & 0.196 \AA & 3.452 & 0.631 e & -0.125 e & -0.471 e \\
           & 2 & 1.203 eV & 0.213 \AA & 0.725 & 0.550 e &  0.197 e & -0.911 e \\
           & 3 & 1.276 eV & 0.674 \AA & 0.000 & 0.651 e &  0.265 e & -1.170 e \\
           & 4 & 0.238 eV & 1.210 \AA & 0.000 & 0.663 e &  0.421 e & -1.201 e \\ \hline
           
        \multirow{5}{*}{Fe} & 0 & -- & 0.035 \AA & 3.406 & 0.307 e & -0.289 e & -- \\
           & 1 & 1.477 eV & 0.051 \AA & 1.607 & 0.337 e & -0.035 e & -0.369 e \\ 
           & 2 & 1.305 eV & 0.235 \AA & 0.003 & 0.377 e &  0.270 e & -0.863 e \\ 
           & 3 & 1.123 eV & 0.801 \AA & 0.000 & 0.489 e &  0.237 e & -1.046 e \\ 
           & 4 & -0.220 eV & 1.310 \AA & 0.000 & 0.578 e & 0.466 e & -1.100 e \\ \hline
           
        \multirow{5}{*}{Co} & 0 & -- & -0.120 \AA & 0.002 & -0.086 e & -0.138 e & -- \\ 
           & 1 & 2.134 eV & -0.002 \AA & 0.001 & 0.164 e & 0.060 e & -0.376 e \\ 
           & 2 & 1.193 eV &  0.274 \AA & 0.000 & 0.262 e & 0.305 e & -0.793 e \\ 
           & 3 & 0.774 eV &  0.887 \AA & 0.000 & 0.401 e & 0.274 e & -0.926 e \\ 
           & 4 & -0.483 eV & 1.256 \AA & 0.000 & 0.505 e & 0.406 e & -1.102 e \\ \hline
           
        \multirow{5}{*}{Ni} & 0 & -- & -0.097 \AA & 0.000 & -0.133 e & -0.076 e & -- \\ 
           & 1 & 1.500 eV & 0.078 \AA & 0.000 & 0.121 e & 0.065 e & -0.286 e \\ 
           & 2 & 0.745 eV & 0.459 \AA & 0.000 & 0.246 e & 0.232 e & -0.588 e \\ 
           & 3 & 0.340 eV & 0.891 \AA & 0.000 & 0.397 e & 0.315 e & -0.848 e \\ 
           & 4 & -0.236 eV & 1.311 \AA & 0.000 & 0.465 e & 0.465 e & -1.068 e \\ \hline
    \end{tabular}
    \caption{Binding energy, outward displacement and magnetic moment in units of unpaired electrons for substituent atoms, M, on the Cu(100) surface with up to 4 CO molecules adsorbed. The change in partial Bader charges of the M atom, the first neighbouring Cu atoms in the surface layer and of the CO molecule.}
    \label{tab:Cu(100)}
\end{table}

\begin{table}[]
    \centering
    \begin{tabular}{|c|c|c|c|c|c|c|c|c|} \hline
        M & \#CO & $E_\mathrm{b\text{-}n^{th}\,CO/M@Cu(111)}$ & $z_M$ & magmom & $\Delta q_M$ & $\Delta q_\mathrm{Cu, 1st}$ & $\Delta q_\mathrm{CO}$ \\ \hline
        \multirow{5}{*}{Sc} & 0 & -- & 0.399 \AA & 0.000 & 1.301 e & -1.167 e & -- \\ 
           & 1 & 0.722 eV & 0.371 \AA & 0.000 & 1.342 e & -0.723 e & -0.495 e \\ 
           & 2 & 0.638 eV & 0.430 \AA & 0.000 & 1.370 e & -0.615 e & -0.648 e \\ 
           & 3 & 0.743 eV & 0.494 \AA & 0.000 & 1.355 e & -0.374 e & -0.966 e \\ 
           & 4 & 0.392 eV & 0.661 \AA & 0.000 & 1.383 e & -0.013 e & -1.383 e \\ \hline
           
        \multirow{5}{*}{Ti} & 0 & --       & 0.108 \AA & 1.041 & 1.037 e & 
           -0.953 e & -- \\ 
           & 1 & 1.146 eV & 0.164 \AA & 0.000 & 1.160 e & -0.585 e & -0.516 e \\ 
           & 2 & 1.177 eV & 0.158 \AA & 0.000 & 1.203 e & -0.310 e & -0.874 e \\ 
           & 3 & 1.094 eV & 0.322 \AA & 0.000 & 1.205 e & 0.021 e & -1.235 e \\ 
           & 4 & 0.386 eV & 0.721 \AA & 0.000 & 1.297 e & 0.106 e & -1.557 e \\ \hline
           
        \multirow{5}{*}{V}  & 0 & --       & 0.108 \AA & 3.022 & 0.737 e & -0.758 e & -- \\ 
           & 1 & 1.228 eV & 0.152 \AA & 2.842 & 0.830 e & -0.589 e & -0.260 e \\ 
           & 2 & 1.394 eV & 0.127 \AA & 0.251 & 0.973 e & -0.049 e & -0.962 e \\ 
           & 3 & 1.375 eV & 0.390 \AA & 0.000 & 1.017 e & 0.162 e & -1.303 e \\ 
           & 4 & 0.304 eV & 0.883 \AA & 0.000 & 1.107 e & 0.229 e & -1.595 e \\ \hline
           
        \multirow{5}{*}{Cr} & 0 & --       & 0.152 \AA & 4.109 & 0.578 e & -0.634 e & -- \\ 
           & 1 & 0.928 eV & 0.153 \AA & 3.476 & 0.709 e & -0.390 e & -0.372 e \\ 
           & 2 & 1.208 eV & 0.171 \AA & 1.275 & 0.755 e & 0.111 e & -0.938 e \\ 
           & 3 & 1.354 eV & 0.483 \AA & 0.000 & 0.818 e & 0.258 e & -1.280 e \\ 
           & 4 & 0.126 eV & 1.089 \AA & 0.000 & 0.922 e & 0.275 e & -1.514 e \\ \hline
           
        \multirow{5}{*}{Mn} & 0 & --       & 0.150 \AA & 4.273 & 0.547 e & -0.606 e & -- \\ 
           & 1 & 0.809 eV & 0.134 \AA & 3.160 & 0.592 e & -0.220 e & -0.472 e \\ 
           & 2 & 1.197 eV & 0.203 \AA & 0.492 & 0.527 e & 0.261 e & -0.907 e \\ 
           & 3 & 1.243 eV & 0.603 \AA & 0.000 & 0.603 e & 0.309 e & -1.177 e \\ 
           & 4 & -0.085 eV & 1.319 \AA & 0.000 & 0.718 e & 0.468 e & -1.376 e \\ \hline
           
        \multirow{5}{*}{Fe} & 0 & --       & -0.025 \AA & 2.895 & 0.278 e & -0.341 e & -- \\ 
           & 1 & 1.423 eV & 0.003 \AA & 1.552 & 0.308 e & -0.121 e & -0.369 e \\ 
           & 2 & 1.299 eV & 0.252 \AA & 0.000 & 0.357 e & 0.331 e & -0.845 e \\ 
           & 3 & 1.002 eV & 0.681 \AA & 0.000 & 0.485 e & 0.370 e & -1.121 e \\ 
           & 4 & -0.346 eV & 1.457 \AA & 0.000 & 0.567 e & 0.531 e & -1.264 e \\ \hline
           
        \multirow{5}{*}{Co} & 0 & --       & -0.119 \AA & 0.000 & -0.061 e & -0.088 e & -- \\ 
           & 1 & 1.962 eV & 0.024 \AA & 0.000 & 0.118 e & 0.145 e & -0.332 e \\ 
           & 2 & 1.110 eV & 0.328 \AA & 0.000 & 0.256 e & 0.361 e & -0.773 e \\ 
           & 3 & 0.629 eV & 0.736 \AA & 0.000 & 0.399 e & 0.496 e & -1.027 e \\ 
           & 4 & -0.568 eV & 1.538 \AA & 0.000 & 0.526 e & 0.561 e & -1.212 e \\ \hline
           
        \multirow{4}{*}{Ni} & 0 & --       & -0.092 \AA & 0.000 & -0.114 e & -0.028 e & -- \\ 
           & 1 & 1.381 eV & 0.069 \AA & 0.000 & 0.002 e & -0.057 e & -0.190 e \\ 
           & 2 & 0.673 eV & 0.320 \AA & 0.000 & 0.239 e & 0.369 e & -0.737 e \\ 
           & 3 & 0.268 eV & 0.612 \AA & 0.000 & 0.368 e & 0.628 e & -1.067 e \\ \hline
           
    \end{tabular}
    \caption{Binding energy, outward displacement and magnetic moment in units of unpaired electrons for substituent atoms M on the Cu(111) surface with up to 4 CO molecules adsorbed. The change in partial Bader charges of the M atom, the first neighbouring Cu atoms in the surface layer and of the CO molecule.}
    \label{tab:Cu(111)}
\end{table}
